\documentclass[8.5pt,twoside,twocolumn]{article}
\usepackage[super,sort&compress,comma]{natbib} 
\usepackage{mhchem}
 \usepackage{times}
\usepackage{sectsty}
\usepackage{balance} 

\usepackage{amssymb, amsmath}
\usepackage{graphicx} 
\usepackage{lastpage}
\usepackage[format=plain,justification=raggedright,singlelinecheck=false,font=small,labelfont=bf,labelsep=space]{caption} 
\usepackage{fancyhdr}
\usepackage{graphicx, latexsym, verbatim}
\usepackage[ansinew]{inputenc}
\usepackage{color}
\usepackage[english]{babel}

\begin{document}

\thispagestyle{plain}
\fancypagestyle{plain}{
\renewcommand{\headrulewidth}{1pt}}
\renewcommand{\thefootnote}{\fnsymbol{footnote}}
\renewcommand\footnoterule{\vspace*{1pt}%
\hrule width 3.4in height 0.4pt \vspace*{5pt}} 
\setcounter{secnumdepth}{5}

\makeatletter 
\def\subsubsection{\@startsection{subsubsection}{3}{10pt}{-1.25ex plus -1ex minus -.1ex}{0ex plus 0ex}{\normalsize\bf}} 
\def\paragraph{\@startsection{paragraph}{4}{10pt}{-1.25ex plus -1ex minus -.1ex}{0ex plus 0ex}{\normalsize\textit}} 
\renewcommand\@biblabel[1]{#1}            
\renewcommand\@makefntext[1]%
{\noindent\makebox[0pt][r]{\@thefnmark\,}#1}
\makeatother 
\renewcommand{\figurename}{\small{Fig.}~}
\sectionfont{\large}
\subsectionfont{\normalsize} 

\fancyfoot{}
\fancyfoot[RO]{\footnotesize{\sffamily{1--\pageref{LastPage} ~\textbar  \hspace{2pt}\thepage}}}
\fancyfoot[LE]{\footnotesize{\sffamily{\thepage~\textbar\hspace{3.45cm} 1--\pageref{LastPage}}}}
\fancyhead{}
\renewcommand{\headrulewidth}{1pt} 
\renewcommand{\footrulewidth}{1pt}
\setlength{\arrayrulewidth}{1pt}
\setlength{\columnsep}{6.5mm}
\setlength\bibsep{1pt}

\twocolumn[
  \begin{@twocolumnfalse}
\noindent\LARGE{\textbf Graphical prediction of quantum interference-induced transmission nodes in functionalized organic molecules}
\vspace{0.6cm}

\noindent\large{\textbf{Troels Markussen,\textit{$^{a}$} Robert Stadler,\textit{$^{b}$} and
Kristian S. Thygesen$^{\ast}$\textit{$^{a}$}}}\vspace{0.5cm}

\noindent\textit{\small{\textbf{Received Xth XXXXXXXXXX 2011, Accepted Xth XXXXXXXXX 2011\newline
First published on the web Xth XXXXXXXXXX 2011}}}

\noindent \textbf{\small{DOI: 10.1039/b000000x}}
\vspace{0.6cm}

\noindent \normalsize{Quantum interference (QI) in molecular transport junctions can lead to dramatic reductions of the electron transmission at certain energies. In a recent work [Markussen \textit{et al., Nano Lett.} 2010, \textbf{10}, 4260] we showed how the presence of such transmission nodes near the Fermi energy can be predicted solely from the structure of a conjugated molecule when the energies of the atomic $p_z$ orbitals do not vary too much. Here we relax the assumption of equal on-site energies and generalize the graphical scheme to molecules containing different atomic species. We use this diagrammatic scheme together with tight-binding and density functional theory calculations to investigate QI in linear molecular chains and aromatic molecules with different side groups. For the molecular chains we find a linear relation between the position of the transmission nodes and the side group $\pi$ orbital energy. In contrast, the transmission functions of functionalized aromatic molecules generally display a rather complex nodal structure due to the interplay between molecular topology and the energy of the side group orbital. }
\vspace{0.5cm}
 \end{@twocolumnfalse}
  ]

\section{Introduction}



\footnotetext{\textit{$^{a}$~Center for Atomic-scale Materials Design (CAMD), Department of Physics,
Technical University of Denmark, DK-2800 Kgs. Lyngby, Denmark; E-mail: trma@fysik.dtu.dk}}
\footnotetext{\textit{$^{b}$~Department of Physical Chemistry, University of Vienna, Sensengasse
8/7, A-1090 Vienna, Austria}}



Quantum interference (QI) effects in molecular junctions has recently been
suggested as an enabling tool for the implementation of molecular
switches~\cite{Baer2002,DijkOrgLett2006, MarkussenJCP2010}, logic
gates~\cite{StadlerNanotech2004}, data storage elements~\cite{Stadler2003} and
thermoelectric devices~\cite{Bergfield2009,Finch2009,Bergfield2010} in 
molecular electronics. These concepts originate from mesoscopics, where electron
transport through waveguide devices has been investigated
already two decades ago~\cite{Sols1989,Porod1992,Porod1993,Debray2000}. In the
context of single-molecule devices, QI were found to be responsible for the observed
reduction of the conductance of a benzene contacted in the meta configuration as
compared to the para and ortho configurations~\cite{Patoux1997,Mayor2003}, and these
findings were rationalized by a variety of different physical pictures, such as
phase shifts of transmission channels or interfering spatial
pathways\cite{Sautet1988,Yoshizawa2008,Fowler2009,Hansen2009}. More recently, the
interest in QI has widened to aromatic molecules of increasing
size~\cite{Walter2004,Cardamone2006,Stafford2007,Ke2008} and also to incoherent
transport in the Coulomb blockade regime~\cite{Hettler2003,Donarini2009}. One way to
induce QI in molecular junctions is to control the electron transmission through
chemical/conformational modification of side groups to aromatic
molecules~\cite{Stadler2003,StadlerNanotech2004,Stadler2005,Papadopoulos2006,Stadler2009},
but also simpler cross-conjugated molecular wires exhibit QI\cite{Solomon2008} and
are promising candidates for implementing switching and rectifying
behaviour~\cite{Andrews2008}.

Within the phase-coherent regime, electron transport through a molecular
junction is described by the energy dependent elastic transmission
function, $T(E)$. In order to illustrate what is meant by QI in a molecular junction, it is instructive to consider the structure in $T(E)$ as arising from three distinct sources. First, we assume that the density of states in the electrodes is constant (wide band approximation) and that each molecular orbital contributes with an independent channel for electron transport. Under these simplifying assumptions, $T(E)$ is a sum of Lorentzian shaped peaks centered at the energy of the
molecular orbitals (MOs). Next, we relax the wide band approximation. This will introduce additional structure in $T(E)$; in particular, the peaks in $T(E)$ will be shifted and change shape. Finally, we relax the assumption of independent transport channels. The structure introduced in $T(E)$ in this last step is referred to as a QI effect. The impact of QI 
on $T(E)$ depends on the relative energies and shape/symmetry of the MOs, and is in general difficult to predict. 
The most characteristic signature
of QI is the presence of transmission nodes, i.e. destructive interference, in the energy gap
between two adjacent MOs. For obvious reasons such transmission nodes are most interesting when they appear in the gap between the highest occupied molecular orbital (HOMO) and lowest unoccupied molecular orbital (LUMO) in which case the QI can suppress the conductance by several orders of magnitude.

In a recent work, we showed that the presence/absence of QI induced transmission nodes under certain conditions can be derived solely from the topological structure of the molecule using a very simple graphical method\cite{MarkussenNanoLett2010,StadlerNanotech2004}. In particular it was demonstrated that the graphical scheme correctly predicts QI interference in five out of ten different anthraquinone based structures. The graphical scheme is exact for a nearest neighbour tight-binding (TB) model with equal on-site
energies, and therefore is expected to work well for all-carbon conjugated systems. The successful application of the scheme to systems containing heteroatoms, such as anthraquinone with its oxygen side groups, may seem surprising and calls for further analysis.

In this paper we relax the assumption of equal on-site energies. By doing so we can write down equations for the zero points of the transmission function directly from
the graphs leading to a "generalized
graphical scheme". For molecules, where the conjugated $\pi$ system is defined by a
linear carbon chain and a side group (which might contain heteroatoms), the scheme predicts a linear relationship between the energy of the QI induced transmission node and the energy of the $\pi$ orbital of the side group. This behavior is confirmed by DFT calculations for molecules with a variety of different side groups. For aromatic molecules the
situation is more complicated due to the interplay between topological and on-site energy effects. In general
several QI induced minima occur in the
transmission function; their energetic position is given by the roots of polynomials in energy which can be derived from the generalized QI graphs. We explore
the dependence of these transmission nodes on the side group on-site energy for the ten anthraquinone structures of
Ref.~\cite{MarkussenNanoLett2010} using both TB and DFT calculations.

The paper is organized as follows. In Section \ref{graphical-scheme}
we summarize the graphical QI scheme from
Ref.~\cite{MarkussenNanoLett2010} and generalize it to the case of varying on-site energies (heteroatoms). More details on the graphical scheme are given in the Appendix. In Sections \ref{linear} and \ref{aromatic} we present our results for QI in functionalized molecular chains and aromatic molecules, respectively. The summary and conclusions are given in Section \ref{summary}.

\section{Graphical scheme} \label{graphical-scheme} 
In this section we review our graphical approach to quantum interference. Some of the
details in the derivation are further explained in Appendix A.

Within a single-particle picture, the transmission probability of an electron impinging on a molecular junction with an energy $E$ is given by
\begin{equation}
T(E) = \text{Tr}[\mathbf{G} \mathbf{\Gamma}_L \mathbf{G}^{\dagger}\mathbf{\Gamma}_R](E)
\end{equation} 
where
$\mathbf{G}=(E\mathbf{I}-\mathbf{H}_{\text{mol}}-\mathbf{\Sigma}_L-\mathbf{\Sigma}_R)^{-1}$
is the Green function matrix of the contacted molecule, $\mathbf{I}$
is the identity matrix, $\mathbf{\Sigma}_{L/R}$ is the self-energy due
to the left/right lead, and $\mathbf{\Gamma}_{L/R}=
i(\mathbf{\Sigma_{L/R}}-\mathbf{\Sigma}_{L/R}^\dagger)$. Let us assume
that the Hamiltonian describing the molecule is given in terms of a
basis consisting of localized atomic-like orbitals,
$\phi_1,\phi_2,\ldots,\phi_N$, and that only the two orbitals $\phi_1$
and $\phi_N$ couple to the leads. In this case the transmission
reduces to
\begin{equation}
T(E) = \gamma(E)^2|\mathbf{G}_{1N}(E)|^2.
\end{equation}
Often the energy dependence on the lead coupling strength, $\gamma$,
can be neglected. It then follows that the transport properties are
entirely governed by the matrix element $\mathbf{G}_{1N}(E)$. The
latter can be obtained using Cramer's rule
\begin{equation}
  \mathbf{G}_{1N}(E)=\frac{C_{1N}(E\mathbf{I}-\mathbf{H}_{\text{mol}})}{\text{det}(E-\mathbf{H}_{\text{mol}}-\mathbf{\Sigma}_L-\mathbf{\Sigma}_R)}
\end{equation}
where $C_{1N}(E-\mathbf{H}_{\text{mol}})$ is the $(1N)$ co-factor of
$(E\mathbf{I}-\mathbf{H}_{\text{mol}})$ defined as the determinant of
the matrix obtained by removing the 1st row and $N$th column from
$(E\mathbf{I}-\mathbf{H}_{\text{mol}}-\mathbf{\Sigma}_L-\mathbf{\Sigma}_R)$
and multiplying it by $(-1)^{1+N}$. Since we assume that only orbitals
$\phi_1$ and $\phi_N$ couple to the leads, the removal of the 1st row
and $N$th column completely removes $\mathbf{\Sigma}_{L,R}$ in the
co-factor.

In the following we shall focus on
$C_{1N}(E\mathbf{I}-\mathbf{H}_{\text{mol}})$ and represent the
determinant graphically. We use the following notation: A hopping
matrix element $t_{ij}$, $i\neq j$ is represented by a (red) line
connecting site $i$ and site $j$. For simplicity we restrict ourself to nearest
neighbor hopping, but the application of the graphical scheme is not
limited to this case. An on-site element $(\varepsilon_i-E)$ is
represented by a (blue) \textit{on-site loop}. Each on-site loop
contributes a factor $(-1)$ (see Appendix A). We shall set the
on-site energy of the ``back-bone'' carbon atoms to zero,
$\varepsilon_0=0$, but let the on site energy of any side groups,
$\varepsilon_{sg}$, be of an arbitrary value. This is the main
difference between this work and the previous
work\cite{StadlerNanotech2004, MarkussenNanoLett2010}. Also, we will
not limit the discussion to transmission at the Fermi energy, but
consider transmission zeros throughout the energy range.

The generalized graphical scheme is summarized as follows: The
transmission zeros can be determined from the zeros of the co-factor
$C_{1,N}(E-\mathbf{H}_{\text{mol}})$. The terms in the co-factor can
be represented graphically by drawing all possible diagrams according
to the rules: (i) In each diagram the external sites connected to
the electrodes must be connected to each other by a continuous path.
(ii) All internal atomic sites must either have one
ingoing \emph{and} one outgoing path \emph{or} have an on-site loop. (iii) The sign of a diagram is $(-1)^p$ where $p$ is the total number of on-site loops and closed hopping loops. When we add up all diagrams constructed according to the above rules we obtain a polynomial in $E$ whose roots represent the QI induced transmission nodes.

\section{Simple molecular wires} \label{linear}

\begin{figure}[htb!]
\centering
\includegraphics[width=0.8\columnwidth]{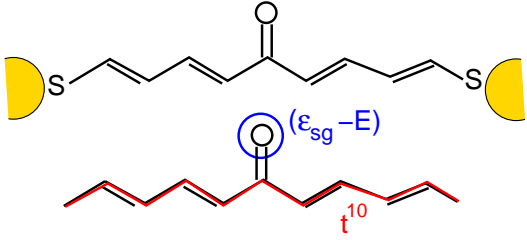}
     \caption{Junction setup for a C9 chain with an oxygen side group (top). The
generalized diagram is shown in the bottom.}
\label{C9-fig}
\end{figure}

We first apply our generalized graphical scheme to the simple system shown in Fig.
\ref{C9-fig} where the $\pi$ system consists of a nine-atom carbon chain (C9) which
is cross-conjugated with a side group. Assuming nearest neighbor hopping there is only one
possible path from left to right which invariably exhibits an on-site loop on the
side group. This loop corresponds to a term $-(\varepsilon_{sg}-E)$ in the
determinant of the cofactor, with $\varepsilon_{sg}$ being the energy of the side
group. The path from the left to right contacts consists of ten hoppings and thus
contributes with a factor $t^{10}$, where $t$ is the nearest neighbor hopping
energy. To find the possible energies at which the transmission is zero, i.e. where
$C_{1N}(\mathbf{H}_{\text{mol}}-E\mathbf{I})=0$, we have only one diagram to consider, which is shown
in the bottom part of Fig. \ref{C9-fig}, and therefore must solve the trivial
equation 
\begin{eqnarray}
-(\varepsilon_{sg}-E)t^{10}=0 \Rightarrow E = \varepsilon_{sg}, \label{eq:liner-zeros}
\end{eqnarray}
for the calculation of the energy, where the transmission zero in this
case is equal to the on-site energy of the side group.  We note that
the energy of the transmission minimum is independent of the on-site
energies of the carbon atoms in the chain and all hopping parameters
in the molecule. Such a case where the energy dependence of the
transmission minimum is linear with the side group energy will in the
following be denoted as a simple side group transmission node. We note
that such transmission nodes directly caused by the side group bears
much resemblance with the general concept of a Fano
resonance\cite{fano,fano3} originating from the coupling of a localized state (the side group)
with a continuum (extended states in the carbon chain). In Ref.
\cite{fano2} we have analyzed the analogy of the simple
side-group transmission nodes with Fano resonances in more detail.

\begin{figure}[htb!]
\centering
\includegraphics[width=0.99\columnwidth]{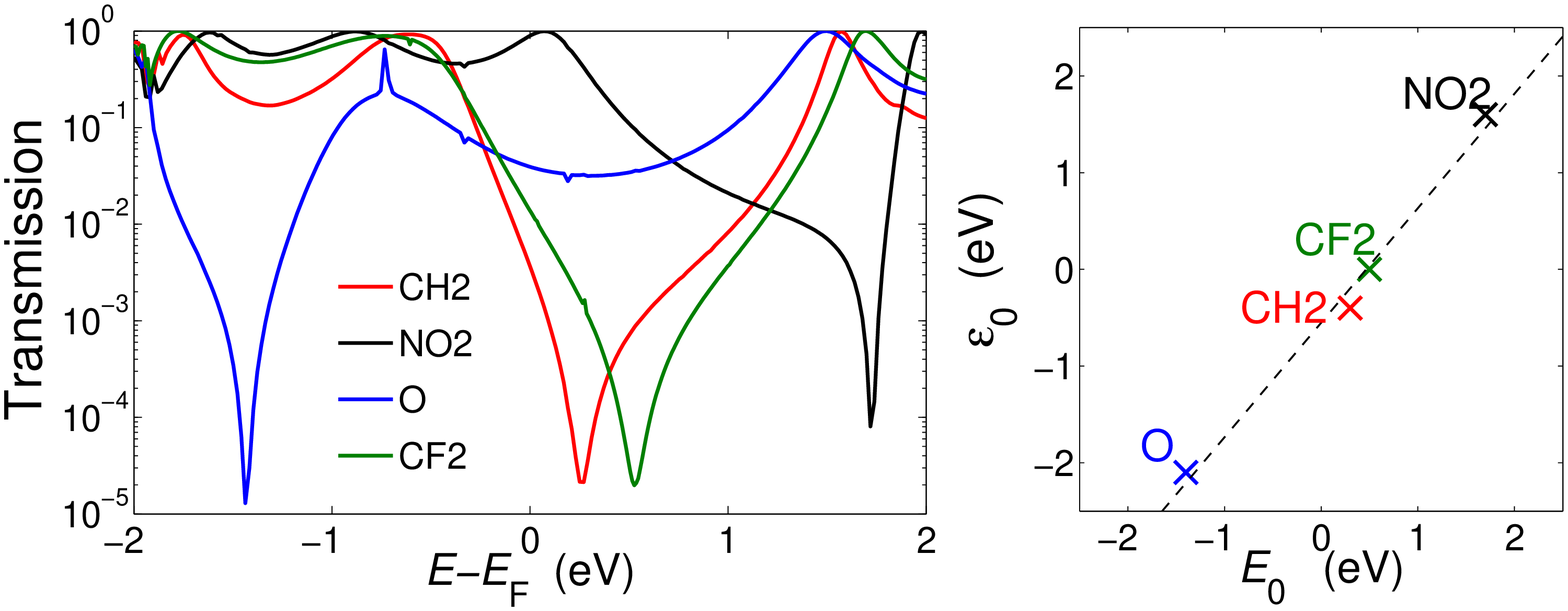}
     \caption{Left: DFT calculated transmission functions for C9 with four different
side groups. Right: $\pi$ state eigenenergy of the side groups,
$\varepsilon_{sg}$, plotted against the energy of the transmission minimum $E_0$.
The side group eigenenergy scales linearly with $E_0$ with a slope of $\sim$1.2. }
\label{C9-trans}
\end{figure}

Figure \ref{C9-trans} (left) shows the transmission for C9-type molecules as
obtained from DFT calculations\cite{calculations}
with four different side groups: CH$_2$, O, NO$_2$,
and CF$_2$. For all four molecules, we observe a distinct transmission minimum,
where its energy can be tuned from $E-E_F=-1.4\,$eV for O to $E-E_F=1.7\,$eV for
NO$_2$. In order to illustrate the validity of the graphical scheme and the
prediction from Eq. \eqref{eq:liner-zeros}, we plot in the right panel of Fig.
\ref{C9-trans} the energy of the side group orbital $\varepsilon_{sg}$ (see Appendix
B for details) vs. the energy of the transmission minima $E_0$. There is a clear linear
dependence (with a slope of ~1.2) in good qualitative agreement with Eq.
\eqref{eq:liner-zeros}. The deviation from a slope of 1 are probably due to the simplifying assumptions in the TB model of only nearest neighbour hopping, only a single $p_z$ orbital on each atom, etc. In the DFT calculations there are interactions beyond nearest neighbours and several orbitals on each atom. In addition to this more subtle charge transfer effects may lead to deviations from the simple TB model. However, overall Eq. \eqref{eq:liner-zeros} gives a good description of the DFT data. 

\section{Aromatic molecules} \label{aromatic}

\begin{figure*}[htb!]
\centering
\includegraphics[width=\textwidth]{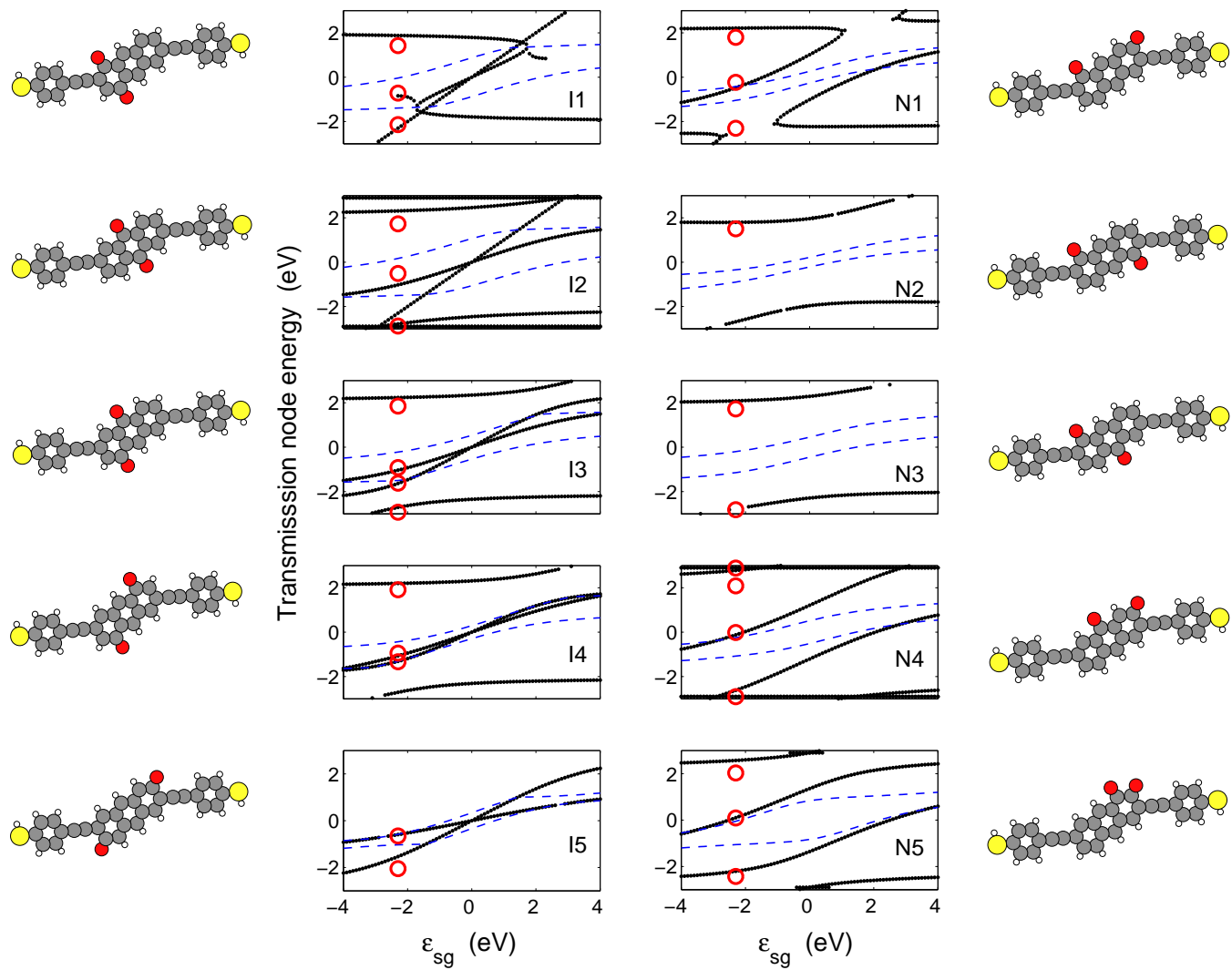}
     \caption{Transmission node energies (black dots) for 10 different anthraquinone
molecules as function of the side group on-site energy, $\varepsilon_{sg}$, as
calculated in a nearest neighbour tight-binding model. The dashed curves show
the HOMO (lower curve) and LUMO (upper curve) energies. The red circles are results from DFT calculations with
oxygen side groups. Each panel in the two central columns corresponds to the
molecular structure in the left- and right most columns. The molecules I1-I5 on
the left side all exhibit QI in terms of transmission nodes within the
HOMO-LUMO gap for most values of $\varepsilon_{sg}$, where no nodes are found
in this energy range for any of the molecules N1-N5 on the right side for any
realistic value of $\varepsilon_{sg}$.}
\label{Is-Ns-fig}
\end{figure*}

We now consider the ten anthracene based molecules illustrated in Fig.
\ref{Is-Ns-fig}, which only differ in the position of the two side groups. In Ref.
\cite{MarkussenNanoLett2010} we studied the same molecules with oxygen side
groups, then they become anthraquinones, and categorized them into two groups, where
five of them exhibited QI in the HOMO-LUMO gap (I1-I5) and the others (N1-N5) did
not. We made this distinction on the basis of our original graphical scheme and our
conclusions were in good agreement with DFT calculations. This prediction, however,
was based on the assumption that the on-site energy of the side group would be
approximately equal to the on-site energy of the carbon $p_z$ orbitals
($\varepsilon_{0}\approx E_F$). From the side group analysis
above, we see that this assumption is questionable for oxygen with a side group
energy of $\varepsilon_{O}\approx-2\,$eV (relative to the Fermi level). We note that this 2 eV difference in on-site energies should be considered relative to the size of the hopping matrix element which is $\sim-3$ eV.  
Still, the success we had in applying the graphical scheme
\cite{MarkussenNanoLett2010,StadlerNanotech2004} to the anthraquinone molecules deserves a more careful analysis and explanation.

For this purpose we make use of a single-orbital nearest neighbour TB model with
carbon on-site energies $\varepsilon_0=0$ and hopping elements $t=-2.9\,$eV. We do not explicitly model the sulfur end group in the TB model, but include it as an effective part of the electrodes. Within
this model we vary only the on-site energy of the side group and compute the
transmission function within a wide band approximation, where the results for the
transmission zeros are shown in Fig. \ref{Is-Ns-fig} as black dots. The lower and
upper dashed lines show the energy of the HOMO and LUMO orbitals, respectively,
while the red circles represent the results of DFT calculations with oxygen
side groups\cite{explain-shift}. First, we note that the simple TB model is in excellent agreement with the DFT calculations, where both result in the same number and the
same energetic positions of transmission zeros. Our second observation is that the
complexity of the dependence of the transmission minima on the on-site energy is
quite striking, in particular when compared to the simple linear scaling found in
the last section for the C9-type molecules. 

\begin{figure}[htb!]
\centering
\includegraphics[width=0.6\columnwidth]{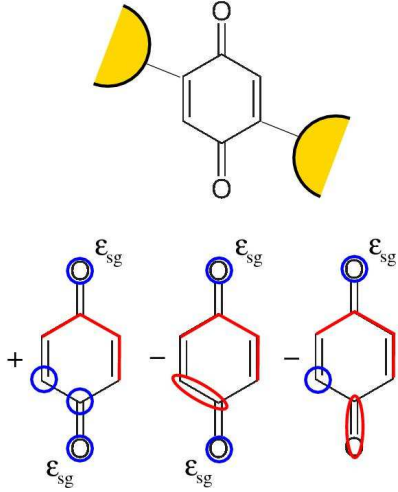}
     \caption{Junction setup for a benzoquinone (top) and all the generalized
diagrams (bottom).}
\label{benzoquinone-fig}
\end{figure}

In order to understand this complexity, we make use of the generalized graphical
scheme introduced above. However, before turning to the anthraquinones we consider the related but simpler
benzoquinone structure shown in the upper panel of Fig. \ref{benzoquinone-fig}. The lower panel shows the generalized diagrams. In addition to the displayed diagrams there are three additional but equivalent diagrams where the external sites are connected via the lower part of the molecule. These diagrams will simply contribute a factor of two to the
resulting polynomials. As for C9 the on-site loops on the side groups again correspond to a
term $-(\varepsilon_{sq}-E)$ in the determinant while for the carbon atoms we define
$-(\varepsilon_0-E) = E$ with $\varepsilon_0=0$. 

Converting the diagrams into a polynomial and taking out their common
factor $(\varepsilon_{sg}-E)t^3$, we arrive at the equation
\begin{eqnarray}
 (\varepsilon_{sg}-E)t^3\left[E^2(\varepsilon_{sg}-E)-t^2(\varepsilon_{sg}-E)
+t^2E\right]=0, \label{benzoquinone-full-eq}
\end{eqnarray}
which determines the transmission zeros. We note that it follows directly from the diagrams that there will be a maximum of four
transmission nodes since there can be a maximum of four on-site loops no matter
how the diagrams are drawn.
The common factor in the equation corresponds to the common part of all three
diagrams, where the upper oxygen on-site loop contributes $(\varepsilon_{sg}-E)$ and
the path from left to right yields $t^3$. We thus immediately see that for
benzoquinone one of the transmission zeros is always at energy
$E=\varepsilon_{sg}$, which is trivially defined by the side group on-site energy as
for the C9 molecule in Fig. \ref{C9-fig}.

\begin{figure}[htb!]
\centering
\includegraphics[width=0.8\columnwidth]{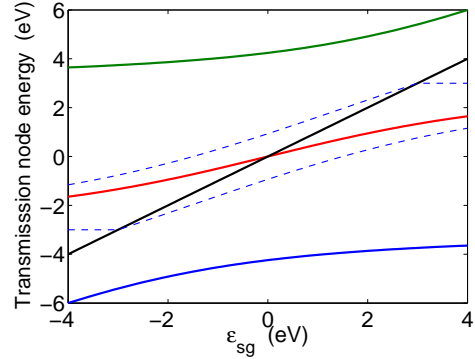}
     \caption{Energy of transmission nodes $E_i$ for the benzoquinone as function of
side group on-site energy, $\varepsilon_{sg}$. The dashed lines mark the HOMO
(lower line) and LUMO (upper line) of the molecule. While the simple side group
transmission node (black) falls outside the HOMO-LUMO gap for large values of
$|\varepsilon_{sg}|$, another node (red curve) stays in the HOMO-LUMO gap for
all values of $\varepsilon_{sg}$.}
\label{benzoquinone-zeros}
\end{figure}

In addition to this simple side group transmission node, Eq.
\eqref{benzoquinone-full-eq} has additional zeros defined by the third order
equation
 \begin{eqnarray}
E^3-\varepsilon_{sg}E^2-2t^2E+t^2\varepsilon_{sg} &=& 0\label{benzoquinone-reduced-eq}.
\end{eqnarray}
While Eq. \eqref{benzoquinone-reduced-eq} can still be solved analytically, this
results in rather lengthy expressions for the energies where $T=0$ as functions of
$\varepsilon_{sg}$, which do not lend themselves to a transparent physical
interpretation. Instead we solve the problem numerically and plot the results for all
four transmission zeros in Fig. \ref{benzoquinone-zeros}. The linear curve
(black) corresponds to the simple side group transmission node at
$E=\varepsilon_{sg}$. The remaining three zeros exhibit a more complicated behavior,
where particularly the red curve is of interest because it is situated
within the HOMO-LUMO gap (marked by the two dashed curves) for the whole range of
$\varepsilon_{sg}$ shown in Fig. \ref{benzoquinone-zeros}. It is striking that although the benzoquinone is much
simpler than the anthraquinone I2 (in terms of the total number of atoms), the nodal
structures of the two corresponding transmission functions are very similar. Most
importantly, this applies for both, the simple side group transmission node (black
curve in Fig. \ref{benzoquinone-zeros}) and the mimimum which is situated within the
HOMO-LUMO gap (red curve in Fig. \ref{benzoquinone-zeros}). Also the remaining two
transmission nodes below the HOMO and above the LUMO are similar for I2 and
benzoquinone. 

\begin{figure}[htb!]
\centering
\vspace{0.5cm}
\includegraphics[width=0.56\columnwidth]{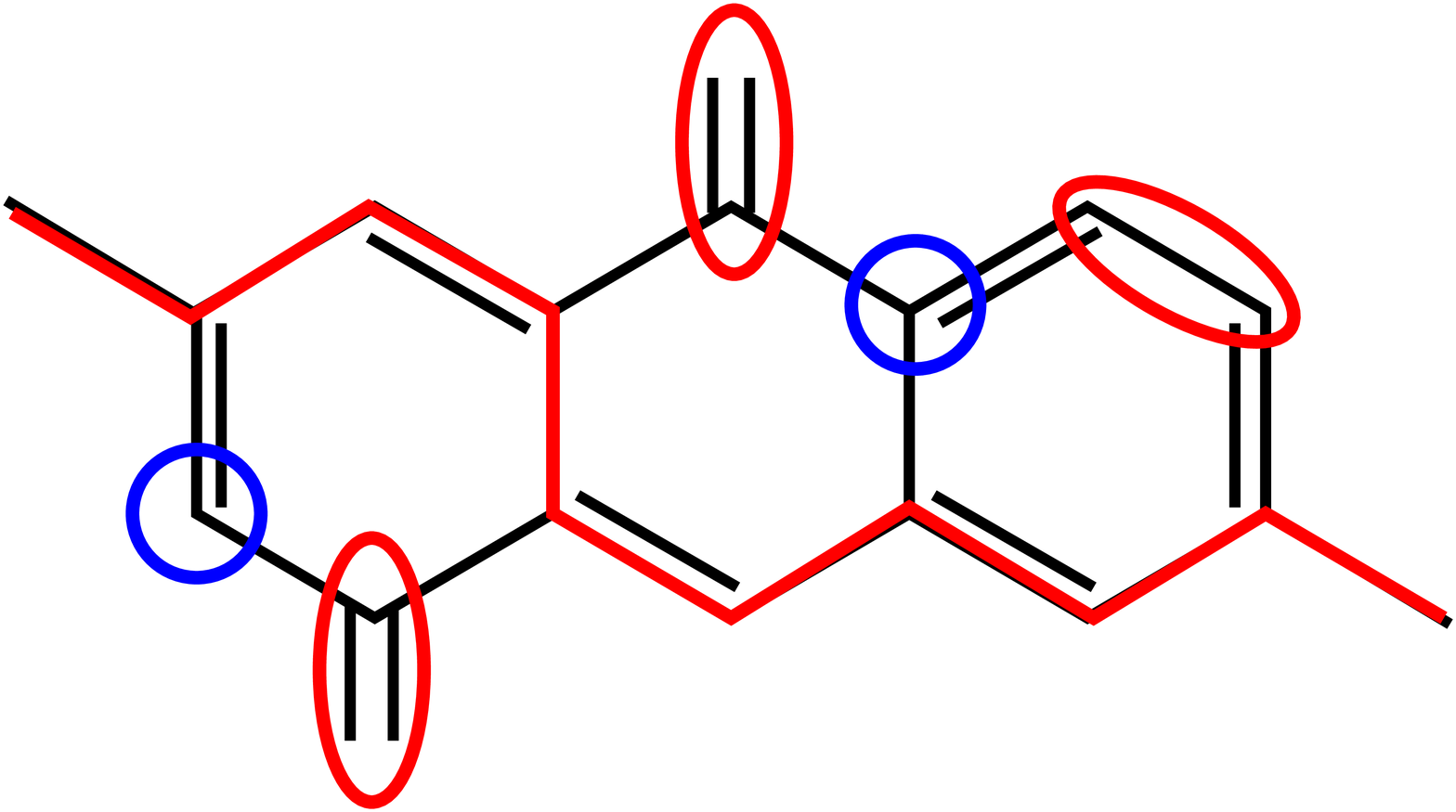}
\put(0,50){I3}

\vspace{0.5cm}

\includegraphics[width=0.56\columnwidth]{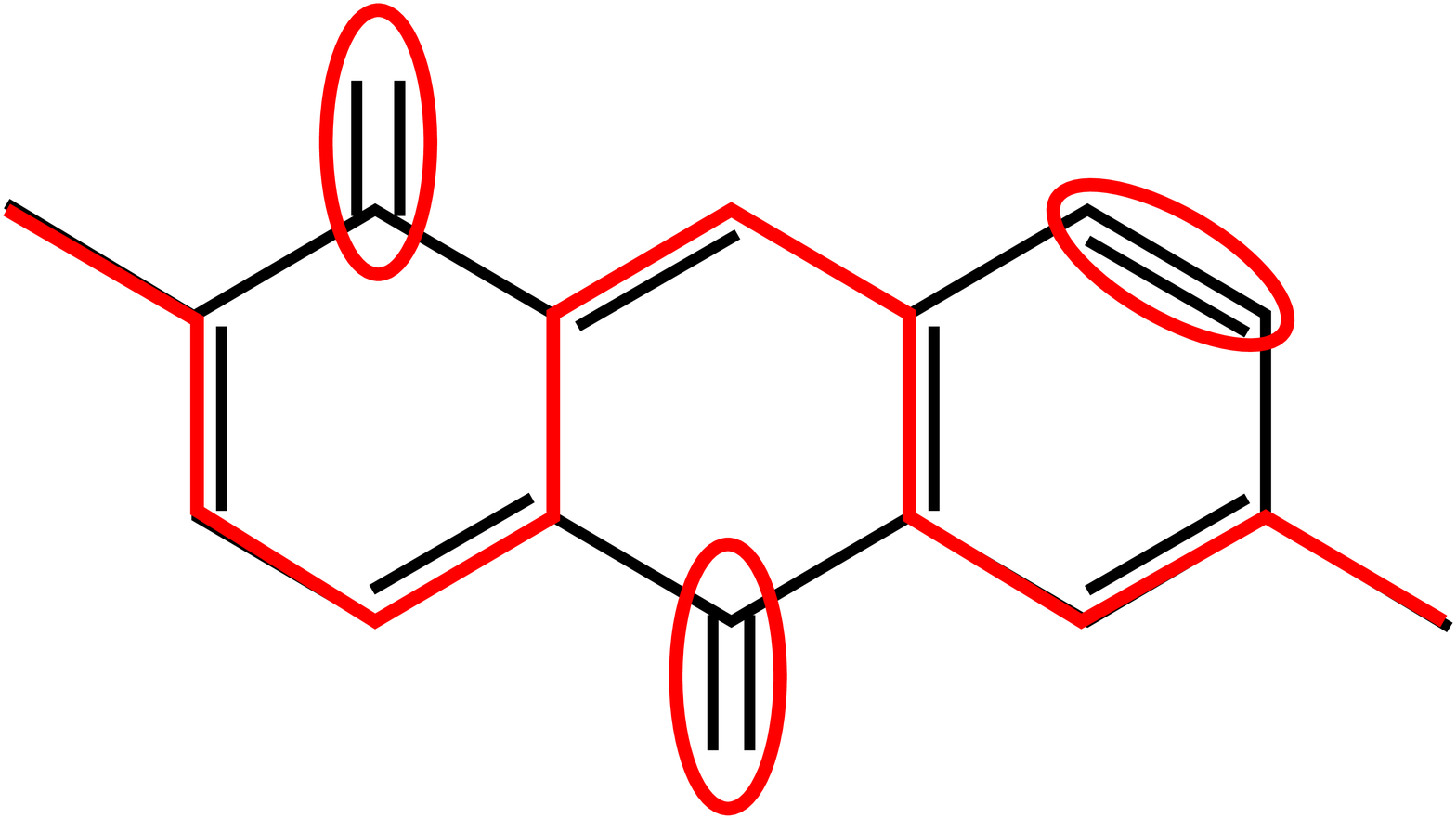}
\put(0,50){N3}
     \caption{Examples of diagrams for I3 (top) and N3 (bottom). I3 will always have
on-site loops (blue circles), but not necessarilly on the side group. The
N3-diagram has no on-site loop and contributes with a finite value at all
energies.}
\label{I3-N3-fig}
\end{figure}

From the five molecules with transmission zeros inside the HOMO-LUMO gap, I1-I5,
only I1 and I2 have a simple side group transmission node. This is straighforward to
explain by using our generalized graphical scheme. I1 and I2 are characterized by
having both side groups on the same benzene ring, and therefore every path from the
left to the right contact has to go through at least one carbon atom with a
side group directly attached. This means that all valid diagrams have at least one
on-site loop, where $(\varepsilon_{sg}-E)$ becomes a common factor among them,
resulting in a simple side group transmission node. For I3-I5 on the other
hand, it is possible to draw valid diagrams with no on-site loop on a side group,
which we illustrate in Fig. \ref{I3-N3-fig} (top) with a diagram for the central
part of I3. If such a diagram is possible, no commom factors can be found for all
diagrams, and hence no simple on-site transmission node exists. The nodal spectra of
the transmission functions for I3-I5 thus has to be solely defined by higher order
polynomial equations which we find indeed in Fig. \ref{Is-Ns-fig}. 

We note that all the
structures I1-I5 have transmission zeros within the HOMO-LUMO gap up to rather large
side group on-site energies $|\varepsilon_{sg}|\lesssim 2.0\,$eV. These are the
molecules expected to show QI based on the predictions from our original graphical
scheme~\cite{MarkussenNanoLett2010,StadlerNanotech2004}. This is consistent with the underlying assumption of that scheme, namely that the variation in on-site energies should be smaller than the hopping matrix element which is $\sim 3$~eV in the present case.

We now turn our attention to the five molecules N1-N5, which we
previously predicted not to have transmission zeros in the relevant
energy range around the Fermi energy~\cite{MarkussenNanoLett2010}.
Strictly speaking our predictions on the basis of the simpler
graphical scheme in Ref.~\cite{MarkussenNanoLett2010} were limited to
the assessment of the absence or occurence of transmission zeros at
the Fermi level, when $\varepsilon_{sg}=0$. From the data plotted in
Fig. \ref{Is-Ns-fig} we can now confirm that even allowing for a
rather large variation of the side group on-site energies, namely for
$|\varepsilon_{sg}|\lesssim2.5\,$eV, there are indeed no transmission
zeros to be found within the HOMO-LUMO gap. We now show how the simple
graphical scheme of Ref.~\cite{MarkussenNanoLett2010} is a special
case of the generalized graphical scheme we introduce in the current
paper. When $\varepsilon_{sg}=\varepsilon_0=0$, all diagrams
containing on-site loops will be zero, which implies for the I1-I5
molecules that all diagrams deliver zero terms to the polynomial
equations and the transmission is zero at $E=0$. Since there are
always two on-site loops, as shown in Fig. 6 (top), the transmission
node at $E=0\,$eV for $\varepsilon_{sg}=0$ will be doubly degenerate.
By changing $\varepsilon_{sg}$ the degenerate nodes split into two and
node crossings are observed for all I1-I5. For the molecules N1-N5
there are, however, always diagrams without on-site loops (an example
is shown for N3 in Fig \ref{I3-N3-fig} (bottom)) which will thus
contribute a finite value to the transmission at $E=0$.

We conclude that the chategorization of the ten antraquinone molecules into
interfering (I1-I5) and non-interfering (N1-N5) predicted by the simple graphical
scheme of Ref.~\cite{MarkussenNanoLett2010} holds for all realistic values of
$\varepsilon_{sg}$. Only for very large values of $|\varepsilon_{sg}|$, which would
require side groups which do not couple to the aromatic $\pi$ system and are
therefore irrelevant for our considerations here, deviations can occur. Fig.
\ref{Is-Ns-fig} shows for instance that I1 has no transmission nodes in the
HOMO-LUMO gap for  $|\varepsilon_{sg}|\gtrsim2.5\,$eV, while for N1 a transmission
node enters the HOMO-LUMO gap for equally rather large $|\varepsilon_{sg}|$ values. 

\section{Conclusions} \label{summary} We have discussed how
characteristic quantum interference (QI) induced nodes in the
transmission function of conjugated molecules can be predicted from
simple graphical considerations involving only the topology of the
molecule and the on-site energy of non-carbon elements of the
$\pi$-system. A previously introduced diagrammatic scheme, strictly
valid for all-carbon molecules, was shown to be qualitatively valid
for molecules containing different atomic species as long as the
on-site energies ($p_z$-orbital energies) do not vary too much
compared to the interatomic hopping strength -- a condition we found
to be met for a range of conjugated molecules with different side
groups (O, CH$_2$, NO$_2$, CF$_2$). We showed that more quantitative
estimates of the transmission node position can be obtained from a
straightforward generalization of the graphical scheme to the case of
finite (and varying) on-site energies. This scheme was then used to
analyze the transmission nodes in linear and aromatic molecules with
side groups. For linear molecular chains a single transmission node
occurs at an energy corresponding to the energy of the side group
$\pi$-orbital while for aromatic molecules the nodal structure of the
transmission function is in general more complex due to a non-trivial
interplay between molecular topology and side group on site energy.
The richness of the nodal structure provides a flexible design tool
for applications of QI in electronic devices based on molecular
junctions.

\textbf{Acknowledgement}  TM and KST acknowledge support from FTP grant nr. 274-08-0408
and The Danish Center for Scientific Computing. The center for Atomic-scale
Materials Design (CAMD) is supported by the Lundbeck Foundation. RS is currently supported by the Austrian Science Fund FWF, projects Nr. P20267 and
Nr. P22548.

\begin{appendix}

\section{Derivation of the generalized graphical scheme}

\begin{figure}[htb!]
\centering
\includegraphics[width=\columnwidth]{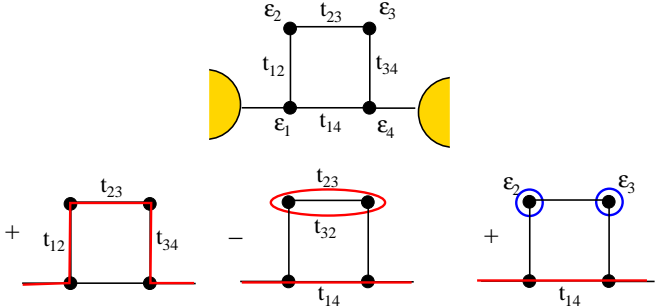}
     \caption{Top: Four site molecule connected to electrodes at site 1 and 4.
Bottom: All the diagrams determining the transmission zeros.} 
\label{4site}
\end{figure}

We shall illustrate the derivation of the graphical scheme by considering the four site
``molecule'' shown in Fig. \ref{4site} (top). The molecule is connected to the
contacts at site 1 and site 4. The Hamiltonian describing the molecule is
\begin{equation}
 \mathbf{H}_{\text{mol}} = \left(
\begin{array}{cccc}
 \varepsilon_1& t_{12} & 0  & t_{14} \\ 
 t_{21} & \varepsilon_2 & t_{23}  & 0 \\ 
 0 & t_{32} & \varepsilon_3 & t_{34} \\ 
 t_{41}& 0 & t_{43} & \varepsilon_4
\end{array}
\right).
\end{equation}
As shown in Section \ref{graphical-scheme}, transmission zeros are determined by the zeros of the co-factor $C_{1N}(E-\mathbf{H}_{\text{mol}})$. The relavant co-factor for the four-site molecule is 
\begin{equation}
 C_{14}(E-\mathbf{H}_{\text{mol}}) = \left|
\begin{array}{ccc} 
 -t_{21} & E-\varepsilon_2 & -t_{23}   \\ 
 0 & -t_{32} & E-\varepsilon_3  \\ 
 -t_{41}& 0 & -t_{43} 
\end{array}
\right|. \label{4-site-cofactor}
\end{equation}

The evaluation of the determinant can be done using Laplace's formula, 
\begin{equation}
\det(A) = \sum_{j=1}^n A_{i,j} (-1)^{i+j} M_{i,j},
\end{equation}
where the minor $M_{i,j}$, is the determinant of the matrix that results from $A$ by
removing the i-th row and the j-th column (i.e. the cofactor without the sign).
Since the on-site energies appears in the first upper diagonal at indices $(i,i+1)$,
each factor $(E-\varepsilon_i)$ have a sign factor $(-1)^{2i+1}=-1$, and thus
contribute with a minus sign. In the minor $M_{i,i+1}$ the on-site terms will still
be in the first upper diagonal and thus contribute with an additional minus sign. 

Writing down the elements of the co-factor gives an equation for the transmission
zeros (the minus sign is included for notational simplicity):
\begin{eqnarray}
-C_{1N}(E-\mathbf{H}_{\text{mol}}) &=&  0 \nonumber \\
t_{12}t_{23}t_{34} - t_{14}t_{23}t_{32} + (\varepsilon_2-E)t_{14}(\varepsilon_3-E) 
&=& 0 .  \nonumber
\end{eqnarray}
The three terms in the co-factor can be represented graphically with the following
convention: Each hopping element $t_{ij}$ is drawn as a line from site $i$ to site
$j$. Note that in all diagrams, the terminating sites (in this case site 1 and 4)
are connected by a continuous path of hopping elements. An on-site term
$(\varepsilon_i-E)$ is drawn as a circle around site $i$ and contributes a
factor (-1). We also note that the loop $t_{23}t_{32}$ going back and forth between
site 2 and 3 in the middlemost diagram gives a minus sign. It is a general rule,
that closed loops, similar to on-site loops, contributs a minus sign. This can
also be derived from Laplace's formula: A hopping element $t_{ij}$ appears at index
$(i-1,j)$ in the co-factor determinant, Eq. \eqref{4-site-cofactor}. By evaluation of
the co-factor along the $(i-1)$'th row the terms involving $t_{ij}$ will contribute
with a sign $(-1)^{i-1+j}$. The new minor appearing after removing the $(i-1)$'th
row and $j$'th column will have the hopping element $t_{ji}$ at index $(j-2,i)$,
assuming without loss of generality that $i>j$. In evaluating this minor along the
$j$'th column, the term $t_{ji}$ comes with a sign $(-1)^{j-2+i}$ and the over all
sign of the combination $t_{ij}t_{ji}$ is $(-1)^{2i+2j-3}=-1$. This shows that a
loop between two neighbouring sites contributs a minus sign. The over all sign
of a diagram is thus $(-1)^p$ where $p$ is the total number of on-site loops and
neighbour loops.

In summary, the transmission zeros can be determined from the zeros of the co-factor
$C_{1N}(E-\mathbf{H}_{\text{mol}})$. The terms in the co-factor can be represented
graphically by drawing all possible diagrams according to the rules: (i) In each
diagram the terminal sites connected to the electrodes must be connected to each
other by a continuous path. (ii) A path can be drawn between sites $i$ and $j$
having non-zero hopping elements, $t_{ij}$. It is not limited to nearest neighbour
interactions only. (iii) All remaining, internal sites must either have one ingoing
\textit{and} one outgoing path \textit{or} have an on-site loop. (iv) The sign of a
diagram is $(-1)^p$ where $p$ is the total number of on-site loops and closed
hopping loops.

\section{Computation of on-site energies}
\begin{figure}[htb!]
\centering
\includegraphics[width=\columnwidth]{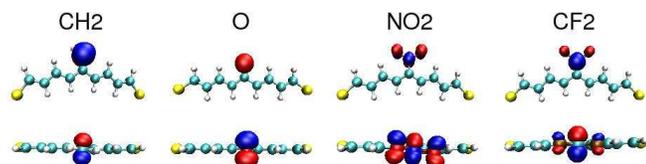}
        \caption{Side group orbitals of C9 responsible for the QI effects close to the
Fermi energy.}
        \label{side group-fig-C9} 
\end{figure}

We calculate the side group on-site energy from the full Hamiltonian matrix,
$\mathbf{H}$, describing the molecule and the Au electrodes. We project onto the
subspace spanned by the basis functions of the side groups:
\[
 \mathbf{h}_{sg} = \mathbf{P}_{sg}\mathbf{H}\mathbf{P}_{sg},
\]
where $\mathbf{P}_{sg}$ has diagonal elements on the indices of the side group basis
functions, and zeros elsewhere. Similarly we get a side group overlap matrix,
$\mathbf{s}_{sg}$. We then diagonalize $s_{sg}^{-1}\mathbf{h}_{sg}$ to find the
side group energies and eigensstates. The corresponding side group orbitals for the C9
molecules responsible for the QI effects are plotted in Fig.
\ref{side group-fig-C9}.

\end{appendix}


\footnotesize{

\begin{mcitethebibliography}{37}
\bibitem[cal()]{calculations}
We calculate the electronic conductance of the molecular contacts using a well
  established combination of DFT and Green's function
  methods\cite{gpaw-review}. The DFT and transport calculations are performed
  using the the real space projector augmented wave method GPAW~\cite{gpaw}
  with the PBE exchange-correlation functional~\cite{PBE}. We initially relax
  the molecule and the two closest Au layers using a double-zeta with
  polarization (DZP) atomic orbital basis set~\cite{gpaw-lcao}. For all the
  molecular junctions considered we find that the anchoring S atom prefers to bind
  to the Au(111) surface in a bridge site slightly shifted toward the hollow
  site. The transmission function of the relaxed junction is calculated using
  the DZP basis set following the standard DFT-Landauer approach described in
  e.g. Ref.~\cite{strange08}\relax
\mciteBstWouldAddEndPuncttrue
\mciteSetBstMidEndSepPunct{\mcitedefaultmidpunct}
{\mcitedefaultendpunct}{\mcitedefaultseppunct}\relax
\EndOfBibitem
\end{mcitethebibliography}

\providecommand*{\mcitethebibliography}{\thebibliography}
\csname @ifundefined\endcsname{endmcitethebibliography}
{\let\endmcitethebibliography\endthebibliography}{}

}

\end{document}